\documentclass[fleqn,usenatbib]{mnras}

\usepackage{newtxtext,newtxmath}

\usepackage[T1]{fontenc}

\DeclareRobustCommand{\VAN}[3]{#2}
\let\VANthebibliography\thebibliography
\def\thebibliography{\DeclareRobustCommand{\VAN}[3]{##3}\VANthebibliography}

\usepackage{graphicx}	
\usepackage{amsmath}	

\def\pu{Puppis A}
\def\rx{RX~J0822--4300}

\def\srg{\textit{SRG}}
\def\art{ART-XC}
\def\erosita{eROSITA}
\def\int{\textit{INTEGRAL}}
\def\urd29{$\rm{URD_{29}}$}
\def\ms6{$\rm{MS_{6}}$}

\def\flux{erg\,s$^{-1}$\,cm$^{-2}$}

\title[{\it SRG}/ART-XC observations of SNR Puppis A]{Wide-field X-ray observations of the supernova remnant Puppis A with the {\it SRG}/ART-XC telescope}

\author[Krivonos et al.]{
R. Krivonos,$^{1}$\thanks{E-mail: krivonos@cosmos.ru (IKI)} V. Arefiev,$^{1}$  I. Lapshov,$^{1}$ E. Filippova,$^{1}$ R. Burenin,$^{1}$ A. Semena,$^{1}$ S. Grebenev,$^{1}$ \newauthor
S. Sazonov,$^{1}$ A. Shtykovsky,$^{1}$ A. Tkachenko,$^{1}$ and A. Lutovinov$^{1}$\\
$^{1}$Space Research Institute (IKI), Profsoyuznaya 84/32, Moscow 117997, Russia\\
}

\date{Accepted 2021 December 9. Received 2021 December 8; in original form 2021 November 11}

\pubyear{2021}

\begin{document}
\label{firstpage}
\pagerange{\pageref{firstpage}--\pageref{lastpage}}
\maketitle

\begin{abstract}
The \textit{Spectrum-Roentgen-Gamma} (\srg) observatory is currently conducting its 4-year all-sky X-ray survey, started on December 12, 2019. The survey is periodically interrupted for technological operations with the spacecraft. These time intervals are usually used by the {\it Mikhail Pavlinsky\/} \art\ telescope to perform calibrations. In this context, \srg\ carried out scanning observations of the \pu\ supernova remnant (SNR) with the aim to check the imaging performance of \art\ and to optimize the technique of image reconstruction for extended objects. Using the unique imaging capabilities of \art\ and its uniform coverage of the entire \pu\ region, we attempted to investigate the morphology of this SNR at energies $\gtrsim4$~keV, and to search for previously unknown X-ray sources. \pu\ was observed in 2019-2020, conducting $1.5^{\circ}\times1.5^{\circ}$ shallow surveys with an exposure of 36 hours. Additional deep pointed observations of the central part of \pu\ were carried out in 2021 lasted 31 hours to highlight the morphology of the extended emission. The X-ray emission of the \pu\ was significantly detected as an extended structure in the 4--6~keV energy band. The morphology of the emission is in general agreement with that observed in soft X-rays previously. The deep sky image of \pu\ obtained with the \art\ telescope is characterized by a {typical SNR shell rim morphology}, an extended emission and a bright emission knot in the north-eastern part of the supernova shell. Also, four point X-ray sources have been detected, including three objects identified in catalogs, and one newly discovered X-ray emitter.
\end{abstract}




\section{Introduction}

\pu\ is one of the best studied Supernova Remnants (SNRs) in our Galaxy, which shows a strong evidence of the shock-cloud interaction. The distance to \pu\ is estimated to be around $2.2$~kpc based on neutral hydrogen observations \citep{2003MNRAS.345..671R}. At this distance, the $50^{\prime}$ visible diameter of \pu\ corresponds to about 30~pc. Near the geometric center of \pu, a central compact object (CCO) \rx\ is located, which has been identified as a stellar remnant left after the supernova explosion \citep{1996ApJ...465L..43P,1999ApJ...525..959Z}. The observed proper motion of the CCO and optical filaments suggests the age of \pu\ to be $4450\pm750$ years \citep{2012ApJ...755..141B}. Early X-ray observations recognised two bright emission knots inside \pu: the ``bright eastern knot'' (BEK) and another one to the north \citep{1982ApJ...258...22P}. The X-ray spectral studies provided evidence that the expanding SNR shell has interacted with dense molecular clouds in a relatively late phase of its evolution \citep{2005ApJ...635..355H,2010ApJ...714.1725K,2012ApJ...756...49K}, rendering \pu\ the first X-ray-identified example of a cloud-shock interaction. No dense molecular cloud is found to be adjacent to the BEK, suggesting that the molecular clumps have been completely destroyed by the shock \citep{2008A&A...480..439P,2010ApJ...725..585A}.

Since \pu\ is the archetypal SNR interacting with dense molecular clouds, it is expected to be a strong emitter of $\gamma$-rays. Based on seven years of Fermi LAT observations, \citet{2017ApJ...843...90X} revealed a noticeable non-thermal $\gamma$-emission located in northeastern and eastern quadrants of the SNR, which is well correlated with the thermal X-ray and IR spatial morphology rather than with the radio emission. This may indicate that some dense clouds are present in \pu\ providing the target material for interaction with cosmic rays. 


\pu\ is one of the brightest SNRs in X-rays. Different areas of its extended emission have been observed with several orbital X-ray telescopes, including {\it Einstein} \citep{1982ApJ...258...22P}, {\it ROSAT} \citep{1993AdSpR..13l..45A}, {\it Chandra} \citep{2005ApJ...635..355H}, {\it Suzaku} \citep{2008ApJ...676..378H} and {\it XMM-Newton} \citep{2006A&A...454..543H,2010ApJ...714.1725K,2012ApJ...756...49K}. The observed X-ray emission is thermal in origin, mostly originated in the shocked interstellar medium \citep{2005ApJ...635..355H}, also showing evidence for the supernova ejecta. \cite{2013AA...555A...9D} presented the most complete and detailed X-ray view of \pu, confirming that this SNR evolves in an inhomogeneous, probably knotty interstellar medium. The HaloSat CubeSat mission recently provided the first soft X-ray (0.4--7 keV) observation of the entire Vela SNR and Puppis A SNR region with a single pointing and moderate spectral resolution \citep{2020AJ....160...20S}.  After this manuscript has been prepared, \cite{2021arXiv211012220M} presented the large-scale distribution of absorption, plasma temperature and emission lines of \pu, using field-scan data acquired by the \srg/\erosita\ telescope during its calibration and performance verification phase.

Due to its relatively large spatial size, \pu\ is a complicated target for X-ray focusing telescopes, since many pointings with different observational configurations have to be combined, which usually requires non-trivial modeling of the astrophysical background. The $36'$ field of view (FOV) of the {\it Mikhail Pavinsky} \art\ telescope, on board the recently launched \srg\ observatory, is well suited for mapping large parts of the \pu\ region. Moreover, the scanning mode observations enable covering even larger regions of the sky with the uniform exposure. The \art\ working 4--30~keV energy band allows previous observations of \pu\ carried out in the similar (medium X-ray) energy bands to be complemented. 

In this work, we present the first results of a series of wide-field observations of \pu\ with \art\ performed in 2019--2021. The paper is organized as follows. Observations of the \pu\ region with \srg\ are described in Sect.~\ref{sec:obs}. Sect.~\ref{sec:data} briefly outlines the data reduction. Sect.~\ref{sec:skymapping} describes the construction of different sky maps of \pu. The detected point X-ray sources are discussed in Sect.~\ref{sec:sources}. Discussion and summary are presented in Sect.~\ref{sec:summary}. 


\section{\art\ observations of \pu}
\label{sec:obs}

The region of the \pu\ SNR was observed in 2019--2021 (Table~\ref{tab:log}) with the {\it Mikhail Pavlinsky\/} Astronomical Roentgen Telescope -- X-ray Concentrator (\art, \citealt{2021A&A...650A..42P}), one of two X-ray telescopes on board the \srg\ observatory \citep{srg}, launched on July 13, 2019, from the Baikonur Cosmodrome. \srg\ is surveying the sky from its halo orbit near the Sun--Earth L2 point. \art\ is a grazing incidence X-ray telescope containing 7 independent modules with their own X-ray mirror assemblies and focal plane CdTe detectors. \art\ energy coverage of 4--30~keV in hard X-rays complements the soft energy response of the other instrument of the \srg\ observatory, \erosita\ \citep{2021A&A...647A...1P}, which is sensitive in the 0.2--8~keV energy band.

\begin{table}
\caption{List of scanning and pointed observations of \pu\ with \srg/\art.}
\label{tab:log}
\centering
\begin{tabular}{c c c c}
\hline\hline
ObsID & Date & Exposure & Step\\ 
\hline
   70019900100 & 29-11-2019 20:35 (MSK) & 18 hr & $10'$ \\
   00003058001 & 24-11-2020 00:50 (MSK) & 18 hr & $4'$ \\
   121100510XX$^{a}$ & 24-05-2021 00:30 (MSK) & 19 hr & $4'$ \\
   121100520XX$^{a}$ & 24-05-2021 20:15 (MSK) & 12 hr & $4'$ \\
\hline
\end{tabular}
\begin{flushleft}
$^{\rm a}$ Two-digit numbers ``XX'' denote indexes of grid nodes. The full list of observations is available on \srg\ website \url{http://srg.cosmos.ru}
\end{flushleft}
\vspace{3mm}
\end{table}

\begin{figure}
\centerline{
\includegraphics[width=0.95\columnwidth,clip]{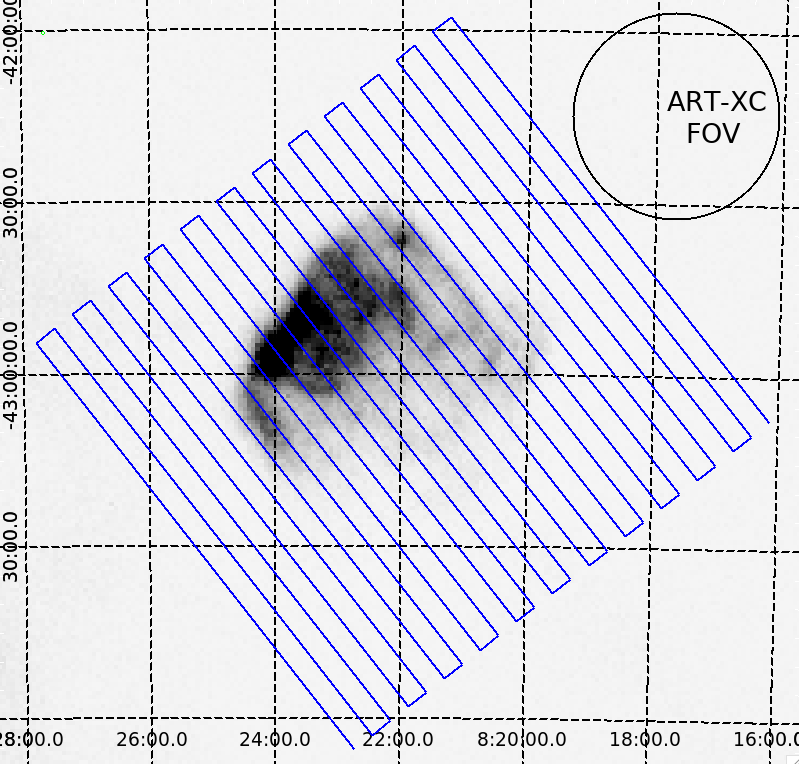}
}
\caption{Scanning mode observations of the \pu\ SNR performed on November 24, 2020 (Galactic coordinates). The scanning step is 4 arcmin. The size of the scanning region, shown as a blue snake-like pattern, is $1.5^{\circ}\times1.5^{\circ}$. The circle indicates the \art\ FOV size ($36'$ in diameter). The image in the background is taken from the {\it ROSAT} All-Sky X-ray Survey (0.4--2.4~keV).}
\label{fig:art_scan}
\end{figure}

\srg\ is currently conducting its planned 4-year all-sky survey\footnote{Observation schedule is available at \url{http://srg.cosmos.ru}}, which is periodically interrupted by technological operations with the spacecraft, mainly associated with orbit corrections. Both the \art\ and \erosita\ telescopes normally perform in-flight on-going calibration before and after an orbital maneuver to check the instrument performance. 

\subsection{Scanning observations}

The first scan pattern (hereafter ``scanning'') observation of the \pu\ region with a total exposure of 18 hrs was carried out during the \srg\ Performance Verification phase in 2019. This scanning observation was done with a scan step of $10^{\prime}$ optimized for the vignetting function of the \erosita\ telescope. 

The second observation with a comparable exposure was performed in 2020 after a planned correction of the \srg\ orbit with the aim to calibrate X-ray imaging of an extended object with an angular size greater than the field of view (FOV) of the \art\ telescope. The scan step size of $4'$ was optimized for the vignetting function of the \art\ X-ray mirrors in order to obtain highly uniform exposure. Fig.~\ref{fig:art_scan} shows the scanning observation, which was carried out on November 24, 2020. The \erosita\ telescope was not operating during this period.

\subsection{Grid observations}

Scanning mode observations allow one to obtain very uniform exposure coverage, but their implementation with the \srg\ platform requires exposure losses in some turning points, referred to as ``parking positions'', which are normally removed from the analysis. The optimal size of the scanning region is found to be from ${\sim}1.5$ to several degrees. In order to cover smaller regions with angular sizes less than ${\sim}1.5$ degrees, the grid of \art\ pointed observations of step $4'$ proves to be more effective. In this case, nearly uniform exposure is concentrated in the central part of the selected sky region.

To cover the brightest part of \pu, we initiated two sets of pointed observations in May 2021 arranged over $6\times6$ (ObsID's 121100510XX) and $5\times5$ (ObsID's 121100520XX) grids with a $4'$ node separation. These grid patterns where shifted by $2'$ with respect to each other, so that the combined data set provides a quasi-uniform exposure of 90--110~ks within ${\sim}24'$ in a diameter. Note that the pointing direction is centered at the maximum of the \art\ vignetting function, which is shifted by ${\sim}2'$ with respect to the FOV center \citep{2021A&A...650A..42P}. The exposure levels of the combined data set in 2021 are shown by black contours in Fig.~\ref{fig:exposure}.



\section{Data analysis}
\label{sec:data}

We reduced raw telemetry data of the \art\ telescope using data analysis tools developed in IKI\footnote{Space Research Institute of the Russian Academy of Sciences, Moscow, Russia}, as briefly described in \citet{2021A&A...650A..42P}. Using the pipeline software (\textsc{artpipeline}), we produced clean calibrated event lists for each of the telescope modules and spacecraft attitude data. The cleaned science data were then processed (\textsc{artproducts}) to obtain exposure and particle background maps as well as sky images.

\begin{figure}
\centerline{
\includegraphics[width=0.98\columnwidth,clip]{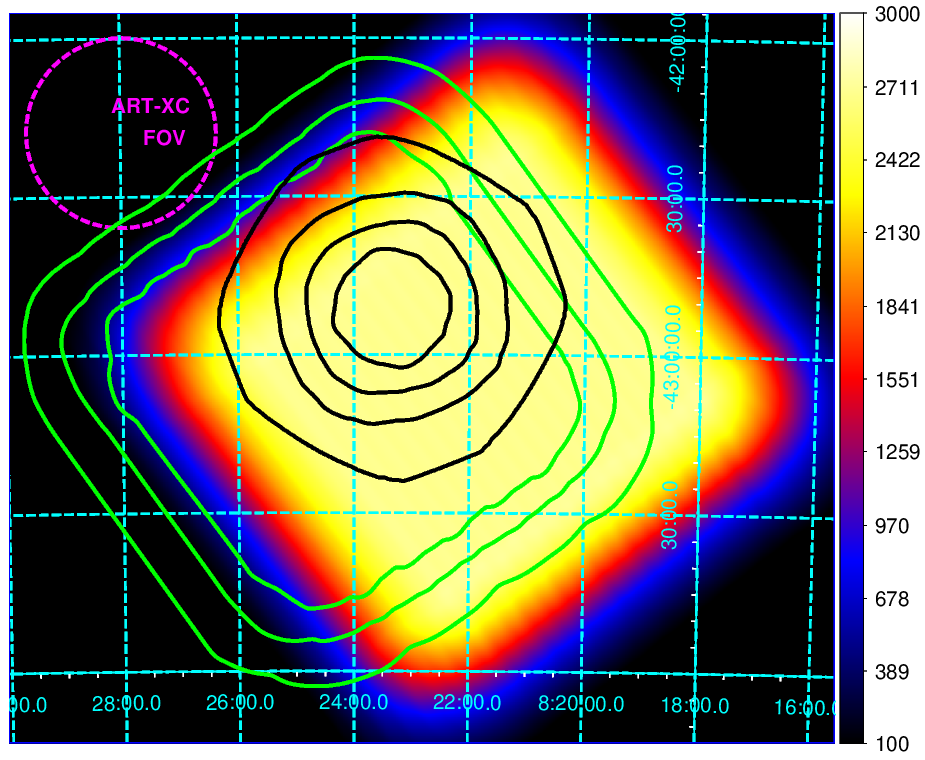}
}
\caption{Exposure map of the \pu\ region scanning mode observation performed in 2020 with a scan step of $4'$ (Table~\ref{tab:log}). The exposure is averaged over all 7 telescope modules. Green contours show 500, 1500 and 2500 second exposure of the scanning mode observations carried out in 2019 with a step of $10'$. Black contours denote exposure levels of 1, 30, 60 and 90~ks accumulated during 2021 pointed observations. Magenta circle demonstrates the \art\ FOV size.}
\label{fig:exposure}
\end{figure}

Fig.~\ref{fig:exposure} shows the exposure map of the \pu\ scanning observations in 2020. It can be seen that the \art\ scanning observations provide a highly uniform coverage of a wide $1.5^{\circ}\times1.5^{\circ}$ region around \pu\ with an exposure of $\sim$2600 seconds in each point (averaged over all 7 telescope modules). A slightly displaced position of the 2019 scanning observations is shown with green contour levels.

\begin{figure*}
\centerline{
\includegraphics[width=0.98\columnwidth,clip]{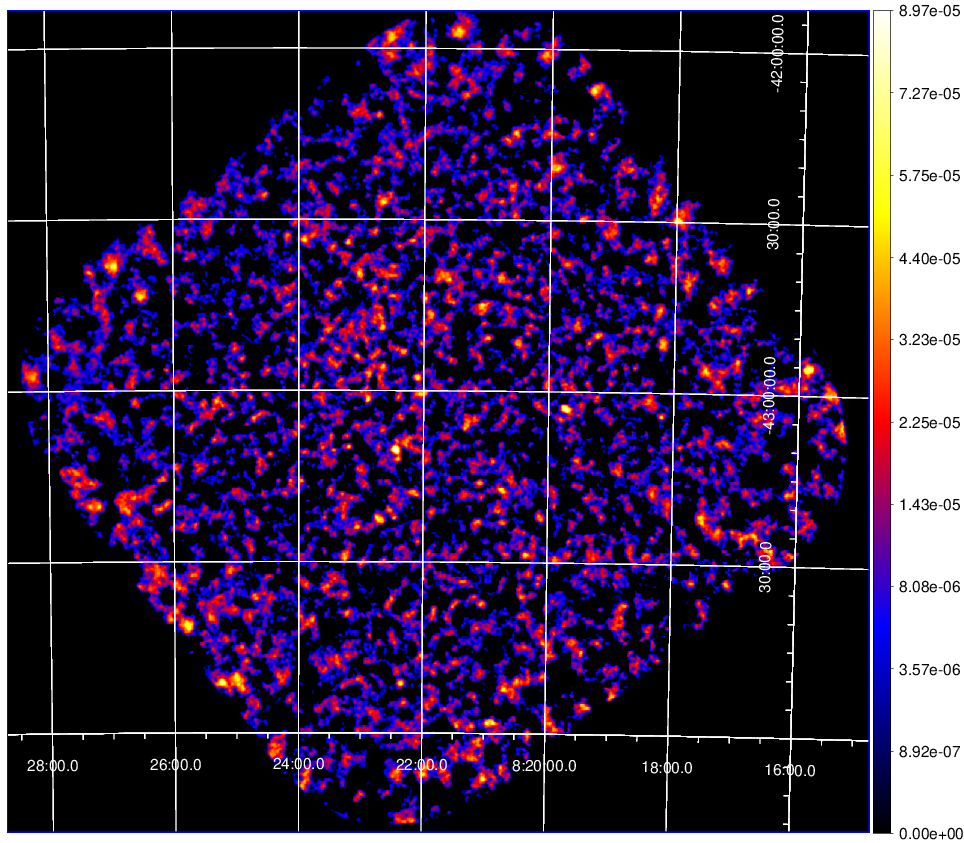}
\includegraphics[width=0.98\columnwidth,clip]{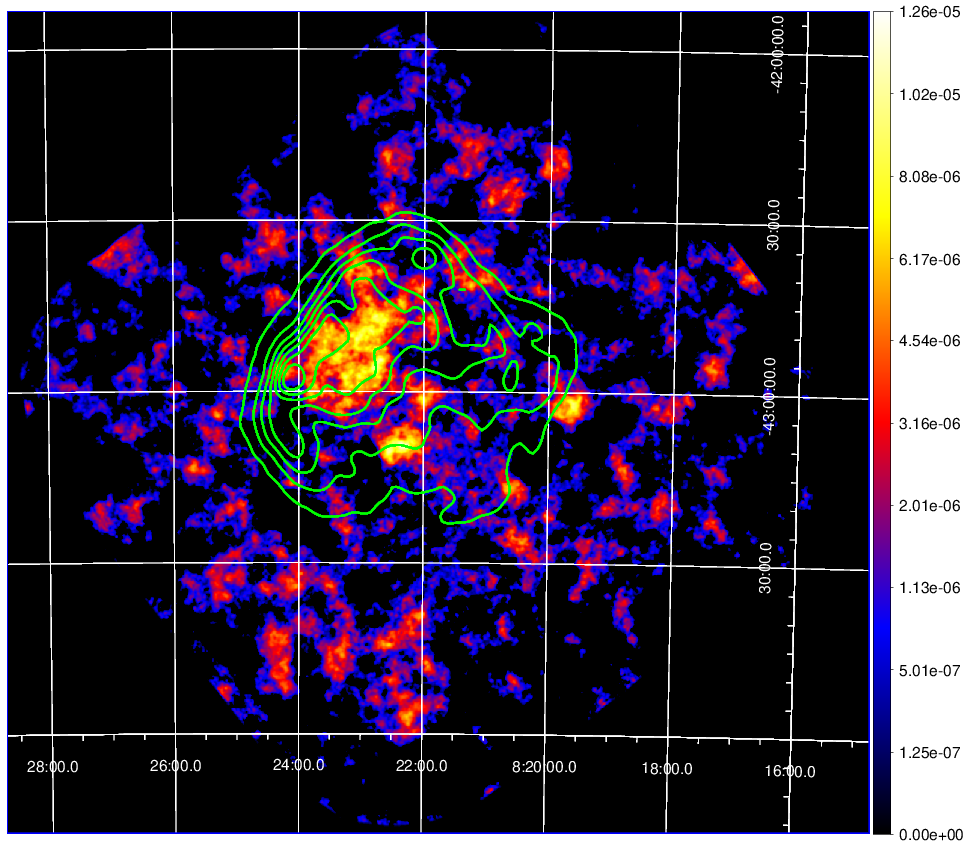}
}
\caption{\art\ image of the \pu\ region in the 4--12~keV (left) and 4--6~keV energy bands (right) obtained in 2020. The images were  smoothed with task \textit{dmimgadapt} from CIAO-4.13 using a tophat kernel. The color bar is in units of counts\,s$^{-1}$\,pixel$^{-1}$, and the angular size of the image pixel is $6.33''$. The white grid indicates Equatorial coordinates in degrees. Green contours demonstrate X-ray surface brightness measured in the ROSAT All-Sky X-ray Survey (corresponds to the background image in Fig.~\ref{fig:art_scan}).}
\label{fig:map}
\end{figure*}

\subsection{X-ray images of \pu}
\label{sec:skymapping}

We combined \art\ cleaned event lists for all 7 telescope modules into sky mosaics in different energy bands from 4 to 12 keV. Figure~\ref{fig:map} (right) shows the sky image of \pu\ in the full 4--12~keV band, which is the most sensitive energy range for detection of point-like X-ray sources, given the energy dependence of the \art\ effective area \citep{2016SPIE.9905E..1JP}. Due to the  sparseness of the photon image, we applied an adaptive smoothing to this image. The image reveals an extended structure of low surface brightness and a number of point-like X-ray sources. We analyzed \art\ sky images of \pu\ in different energy bands from 4 to 12 keV and found that the extended emission of the supernova remnant is most pronounced in the 4--6~keV energy range, as shown in Fig.~\ref{fig:map} (right). At higher energies $\gtrsim6$ keV, we do not detect any significant excess of the \pu\ extended emission above the background. In general, the 4--6~keV emission of \pu\ is consistent with the brightest part of the SNR known from {\it ROSAT} observations \citep{1993AdSpR..13l..45A}.

\begin{figure}
\centerline{
\includegraphics[width=0.98\columnwidth,clip]{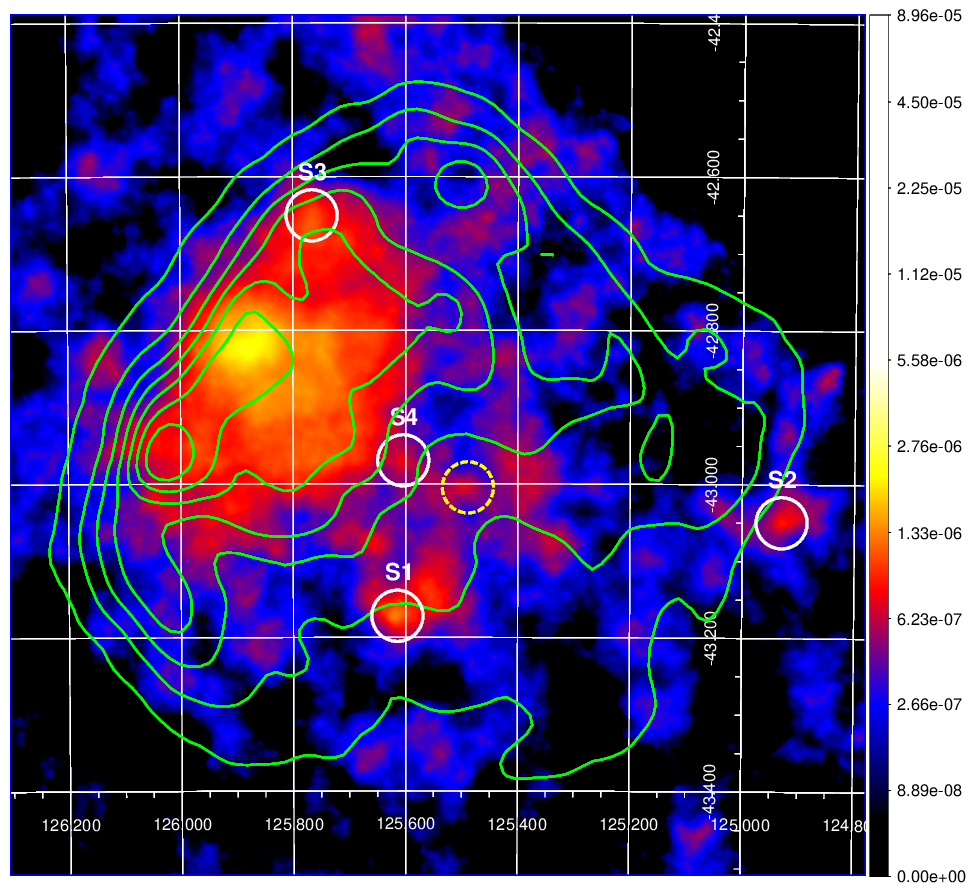}
}
\caption{\art\ 4--6~keV image of the full \pu\ SNR region based on \srg/\art\ data acquired in 2019--2020. The white circles show the  positions of the detected pointed X-ray sources listed in Table~\ref{tab:srclist}. The yellow dashed circle denotes the position of CCO RX~J0822--4300.}
\label{fig:total}
\end{figure}

Figure~\ref{fig:total} shows the combined 2019--2020 image of the full region of the \pu\ SNR.  In order to reveal faint features, the image was processed with a simple low-pass filter with a cutoff frequency ${\sim}0.1$ arcminute$^{-1}$. Among the extended structures consistent with ROSAT contours, the image reveals a bright emission knot in the northeastern part of the SNR centered approximately at RA=08:23:46, Dec=-42:53:38 (J2000). This bright knot is not spatially consistent with the bright eastern knot revealed previously in soft X-rays \citep{1982ApJ...258...22P} and shown by the highest {\it ROSAT} contour level. 

\begin{figure}
\centerline{
\includegraphics[width=0.98\columnwidth,clip]{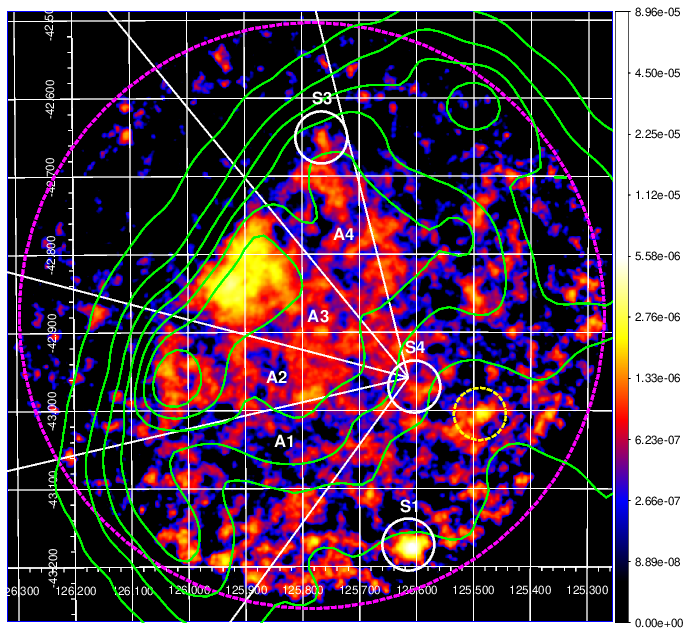}
}
\caption{\art\ 4--6~keV zoomed-in image of the \pu\ region based on the combined \srg/\art\ data set acquired in 2019--2021. The magenta circle with $50'$ in diameter denotes the region with the \art\ exposure greater than 30~ks.}
\label{fig:filtered}
\end{figure}

Figure~\ref{fig:filtered} shows the deepest low-pass filtered image of the central part of the \pu\ SNR obtained with \art\ in the 4--6~keV band using all available observations carried out in 2019--2021. The image reveals four point X-ray sources: S1, S3, S4 and the CCO RX~J0822--4300 (see Sect.~\ref{sec:sources}), as well as the \pu\ diffuse emission. The overall distribution of the SNR 4--6~keV emission follows the {\it ROSAT} soft X-ray contours. As seen from the image, the bright emission knot visible in the 2019--2020 images became more prominent. Using a simple aperture analysis, we estimated the 4--6~keV background-corrected flux of the knot within the $R=4'$ circle as ${\sim}0.1$~mCrab or ${\sim}5.5\times10^{-13}$\flux.

\begin{figure}
\centerline{
\includegraphics[width=0.98\columnwidth,clip]{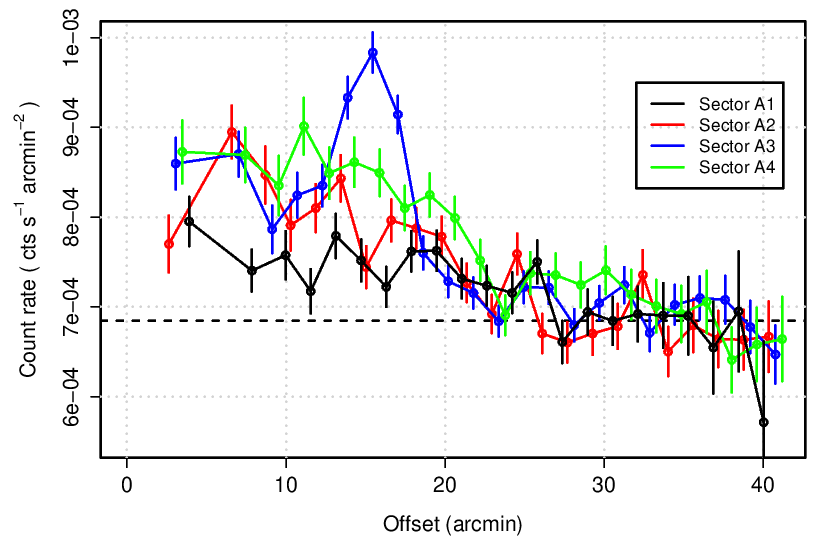}
}
\caption{Radial profile of \pu\ in four sectors defined in Fig.~\ref{fig:filtered}. {Dashed horizontal line represents background level measured at offset $>25'$.}}
\label{fig:prof}
\end{figure}

{To explore the arc-shaped shock front of the SNR, we constructed radial profiles in four sectors centered at the remnant’s optical
expansion center at RA=$08^{\rm h}22^{\rm m}27\fs5$, Dec.=$-42\degr57'29\farcs$  \citep{1988srim.conf...65W,2012ApJ...755..141B}, as shown in Fig.~\ref{fig:filtered}. Sectors A1-4 were chosen to highlight the SNR emission as a function of the angular offset from the center out to the distance where it dims below the background level (Fig.~\ref{fig:prof}). All the profiles demonstrate a drop of the emission at the same angular offset of ${\sim}20'$. Only the radial profile A3 reveals a strong excess of the emission knot in the eastern part of the shell.}





\subsection{Point-like X-ray sources}
\label{sec:sources}

\begin{figure}
\centerline{
\includegraphics[width=0.98\columnwidth,clip]{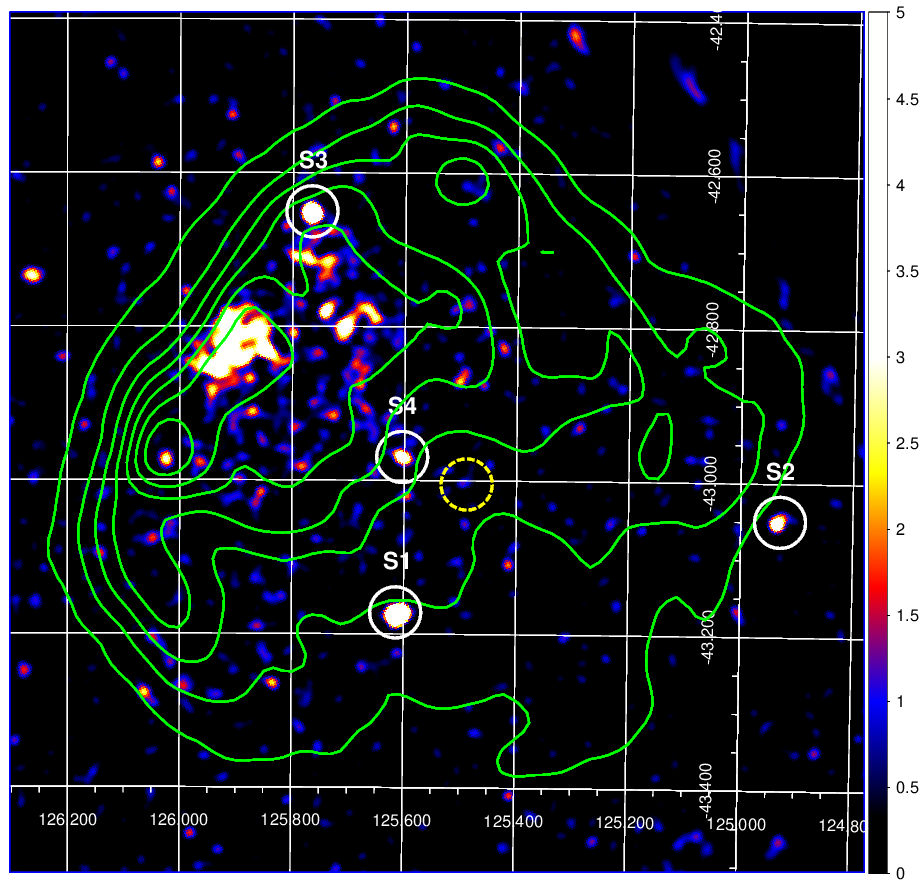}
}
\caption{\art\ 4--12~keV sky image of the full \pu\ SNR region based on \srg/\art\ data acquired in 2019--2021. The image is shown in units of logarithm of the ratio of the observed net count probability under the condition with and without a point-like source at particular sky position, which corresponds to $-log_{10}(1 - P)$, where $P$ is the probability that the source is not related to the background fluctuations. See description of the overplotted visualization in Figs~\ref{fig:map}-\ref{fig:total}.}
\label{fig:andy}
\end{figure}

We performed a search for point X-ray sources on the combined 2019--2021 image of \pu. The source detection was done by the wavelet decomposition method (\texttt{wvdecomp}, \citealt{1998ApJ...502..558V}), along with matched filter and maximum likelihood fitting algorithms. The method uses a probabilistic approach for detection of X-ray sources from non-negative counts following the Poisson statistics \citep[see e.g.,][]{2016ApJ...825..132H}, and taking the \art\ vignetting and point spread function (PSF) into account. Figure~\ref{fig:andy} shows the source detection probability map based on a given estimate of background counts. This map provides a direct indicator of how likely or unlikely it is to have a new point-like source. The map reveals a significant detection of four X-ray sources. In addition, Fig.~\ref{fig:andy} reveals extended emission from \pu, including the aforementioned bright knot. However, it is not properly described by the method since it is optimized only for point-like sources.

\begin{table}
\noindent
\centering
\caption{The list of sources detected in the 4--12~keV band for the combined 2019--2021 data set.}\label{tab:srclist}
\centering
\vspace{1mm}
  \begin{tabular}{ccccccccc}
\hline\hline
Label &SRGA Name & RA$^{\rm a)}$ & Dec$^{\rm a)}$ & Flux$^{\rm b)}$ \\
&     & [ J2000 ] & [ J2000 ] &   \\
\hline
S1 & J082226.3--431020 &  125.6097 & -43.1722 &  $7.4 \pm  0.5$ \\
S2 & J081943.7--430253 &  124.9320 & -43.0480 &  $3.5_{-0.6}^{+0.9}$   \\
S3 & J082303.9--423914 &  125.7660 & -42.6539 &  $2.4\pm0.3$   \\
S4 & J082225.0--425810 &  125.6041 & -42.9695 &  $0.8 \pm  0.2$ \\

\hline
\end{tabular}
\begin{flushleft}
$^{\rm a}$ The positional uncertainty is ${\sim}15''$.\\
$^{\rm b}$ The flux is expressed in units of $10^{-13}$\flux\ and estimated as the aperture one.
\end{flushleft}
\end{table}

Table~\ref{tab:srclist} shows the list of four detected sources with a signal-to-noise ratio greater than ${\sim}5\sigma$. The S1 source is spatially consistent with 4XMM~J082226.8$-$431027. According to the {\it XMM-Newton}  catalog\footnote{\url{http://xmm-catalog.irap.omp.eu}} \citep{2017ApJ...839..125Z}, this source has a relatively hard spectrum with $\Gamma{\sim}1.5$ and a flux of $(8.7\pm0.2)\times10^{-13}$ \flux\ in the 4.5--12~keV band, which agrees well with our \art\ measurement. The S2 source does not have any known X-ray counterpart within $1'$ from the \art\ position. The S3 source, located in the bright rim of the supernova shell, can be identified with 2CXO~J082303.0-423901 from the {\it Chandra} source catalog Release 2.0 \citep{2010ApJS..189...37E}, located ${\sim}15''$ away from the \art\ position. The last source, S4, is consistent with the position of the 4XMM~J082224.9$-$425811 source as well as with its 4.5--12~keV flux of $(9.2 \pm 1.3)\times 10^{-14}$ \flux\, which renders it a likely counterpart. We finally note that for the exposure of 6--7~ks in these \art\ observations, the typical sensitivity is ${\sim}1.5\times10^{-13}$\flux\ (4--12~keV). But it is higher in several regions, which allows us to detect fainter sources, like S4.  

The CCO RX~J0822$-$4300, which emits thermal X-rays, is not detected with \art\ in the full 4--12~keV band. However, it is revealed in the 4--6~keV image (see Fig.~\ref{fig:filtered}) at the position of RA=$08^{\rm h}21^{\rm m}59\fs49$, Dec.=$-43\degr00'16\farcs4$ (uncertainty ${\sim}30''$), which is spatially consistent with the position of the CCO determined by \cite{2012ApJ...755..141B} using {\it Chandra} data. The estimated flux of CCO is $(6.6\pm2.4)\times10^{-14}$ \flux\ in the 4--6~keV energy band.







\section{Discussion and conclusions}
\label{sec:summary}

We have presented wide-field observations of the SNR \pu\ region with the {\it Mikhail Pavlinsky} \art\ telescope on board the \srg\ observatory. The \pu\ supernova remnant has been significantly detected in the 4--6~keV energy band as a spatially extended inhomogeneous emission. In general, the X-ray map of \pu\ obtained by \art\ (see Fig.\,\ref{fig:filtered}) demonstrates the morphology consistent with that observed in previous soft X-ray studies. Among the most noticeable features one can mention a { typical shell rim morphology} of the SNR and the bright emission knot in the northeastern part of the remnant. The  {shell-like morphology} of the SNR is consistent with the {\it ROSAT} contours as well as with later {\it XMM-Newton} \citep{2010ApJ...714.1725K,2013AA...555A...9D} and {\it Chandra} \citep{2005ApJ...635..355H} observations. The overall $4-6$~keV \art\ morphology of \pu\ is consistent with that simultaneously observed by \srg/\erosita\ in $3.00-3.25$~keV band \citep{2021arXiv211012220M}.

The bright extended knot observed by \art\ is located inside the SNR shell and is spatially consistent with the region of highest plasma temperature (${\sim}0.75$~keV) observed with \srg/\erosita\ in \pu\ \citep{2021arXiv211012220M}, which indicates that the $4-6$~keV bright emission knot seen by \art\ belongs to hard tail of high temperature thermal emission. It is interesting to note that this region is also  spatially consistent with the non-thermal $\gamma$-emission detected by {\it Fermi} in the northeastern quadrant of the remnant \citep{2017ApJ...843...90X}, which raises open question of the presence of a non-thermal emission component in \pu\ for future studies.


Among the four X-ray point sources significantly detected in the 4--12~keV band, three are associated with known {\it XMM-Newton} and {\it Chandra} sources and one has been detected in X-rays for the first time. The central compact object RX~J0822--4300 is detected only in the 4--6~keV energy band, with a flux of $(6.6\pm2.4)\times10^{-14}$ \flux.

In the forthcoming paper, we will present a spectral study of \pu\ based on the \srg/\art\ observations reported here.

\section*{Acknowledgements}

The {\it Mikhail Pavlinsky} \art\ telescope is the hard X-ray instrument on board the \srg\ observatory, a flagship astrophysical project of the Russian Federal Space Program realized by the Russian Space Agency in the interests of the Russian Academy of Sciences. The \art\ team thanks the Russian Space Agency, Russian Academy of Sciences, and State Corporation Rosatom for the support of the \srg\ project and \art\ telescope. We thank Lavochkin Association (NPOL) with partners for the creation and operation of the \srg\ spacecraft (Navigator). This work has made use of the ROSAT Data Archive of the Max-Planck-Institut fuer extraterrestrische Physik (MPE) at Garching, Germany. This research has made use of data obtained from the Chandra Source Catalog, provided by the Chandra X-ray Center (CXC) as part of the Chandra Data Archive. RK acknowledges support from the Russian Science Foundation grant 19-12-00369 in working on this article.

\section*{Data Availability}

The data of the \art\ telescope used in this paper, as well as data analysis software is not available for public. However, there are plans to provide public access to the \art\ scientific archive in the future.



\bibliographystyle{mnras}
\bibliography{main} 








\bsp	
\label{lastpage}
\end{document}